**RESEARCH**                                                                                           **Open Access**

# Dual-trigger release of berberine chloride from the gelatin/perfluorohexane core–shell structure

Mahshid Givarian[1*] , Fathollah Moztarzadeh[1], Maryam Ghaffari[1], AmirHossein Bahmanpour[1], Maryam Mollazadeh-Bajestani[2], Manijhe Mokhtari-Dizaji[3] and Fatemeh Mehradnia[4]

## Abstract

**Background**  The development of smart nanocarriers that enable controlled drug release in response to internal and external triggers is an emerging approach for targeted therapy. This study focused on designing pH-sensitive, ultrasound-responsive gelatin/perfluorohexane (PFH) nanodroplets loaded with berberine chloride as a model drug.

**Results**  The nanodroplets were prepared using an emulsion technique and optimized by varying process parameters like homogenization rate, polymer concentration, surfactant, drug, and perfluorocarbon content. The optimal formulation yielded nanodroplets with a particle size of 281.7 nm, a drug encapsulation efficiency of 66.8 ± 1.7%, and a passive drug release of 15.4 ± 0.2% within 24 h. Characterization confirmed successful encapsulation and pH-responsive behavior. Ultrasound stimulation significantly enhanced drug release, with 150 kHz being more effective than 1 MHz in triggering acoustic droplet vaporization while minimizing heat generation. After 10 min of radiation, the optimal formulation showed 89.4% cumulative drug release. The nanodroplets displayed stability over 1 month at 4°C.

**Conclusions**  Overall, the dual-triggered nanodroplets demonstrate excellent potential for controlled delivery and targeted release of berberine chloride.

*Correspondence:
Mahshid Givarian
m.geevarian2@aut.ac.ir
Full list of author information is available at the end of the article

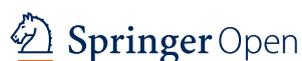

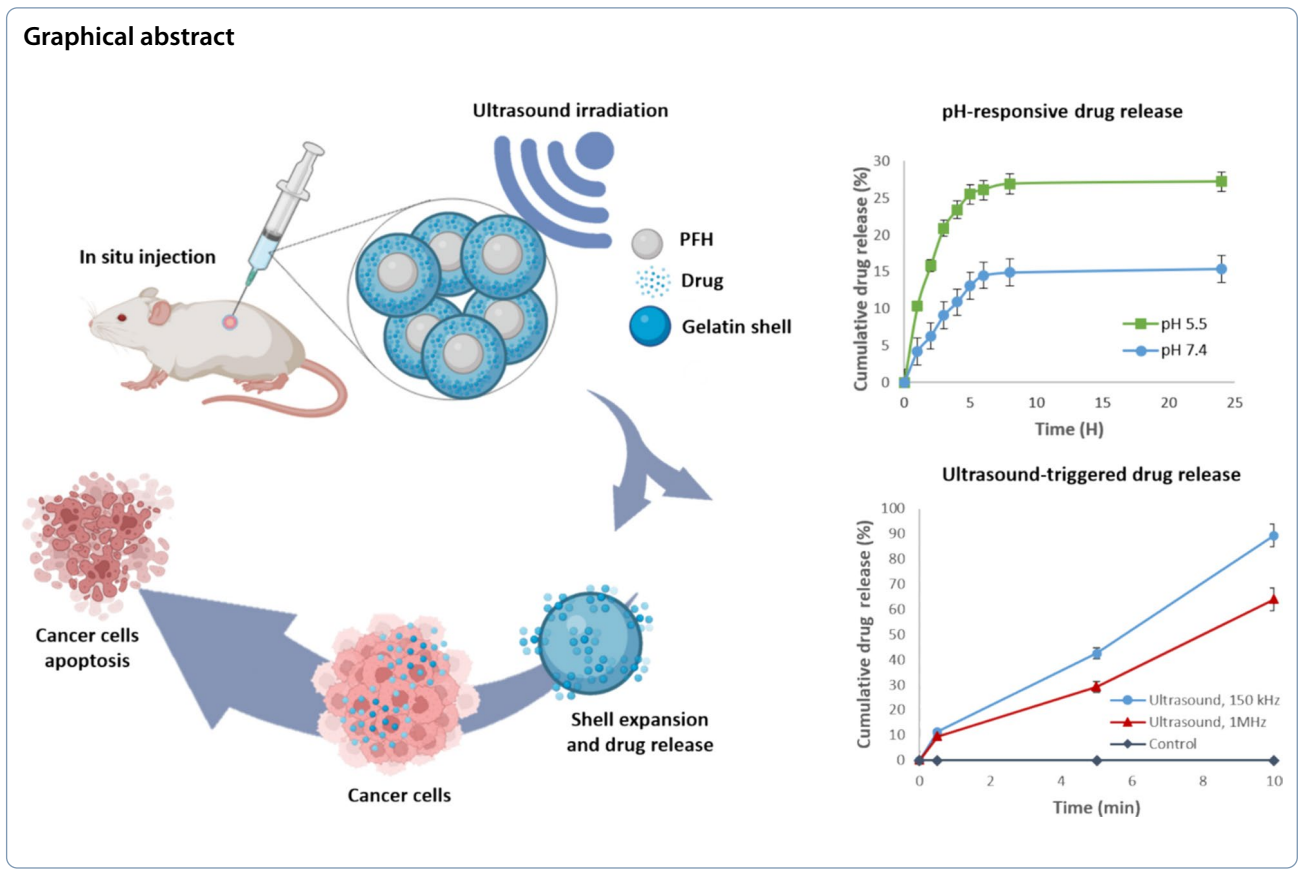

## Background

Recent advances in nanotechnology have revolutionized drug delivery systems by introducing nanocarriers for targeted and controlled drug release. Nanocarriers are nano-sized structures designed to encapsulate and transport therapeutic agents, such as drugs or genes, to specific target sites in the body. These nanocarriers can be composed of various materials, including lipids, polymers, metals, or hybrid combinations (Majumder and Minko 2021; Zhao et al. 2020). The small size and increased surface-to-volume ratio of nanocarriers offer significant advantages in drug delivery. These characteristics enable improved drug penetration, enhanced cellular uptake, and interaction with target cells and tissues, ultimately leading to more effective therapeutic outcomes (Naidu and Paulson 2011).

By harnessing the power of ultrasound energy and responsive materials, drug delivery can now be targeted with precision, resulting in fewer side effects and increased treatment effectiveness. This approach allows for a controlled release of encapsulated therapeutic agents, specifically localized to the target tissue using an external stimulus. Ultrasound not only enables precise targeting but also enhances drug penetration through sonoporation, ensuring better interaction with cells and tissues. The integration of ultrasound and nanotechnology provides a noninvasive and patient-friendly approach, improving the quality of life for individuals undergoing treatment and opening up new possibilities for personalized medicine (Moradi Kashkooli et al. 2023). Liquid perfluorocarbon (PFC) nanoemulsions are fluorinated hydrocarbons that are chemically inert, biocompatible, and have excellent oxygen and gas-carrying capacity. When formulated into nanoemulsions, PFCs can serve as carriers for therapeutic agents and exhibit unique properties that can be triggered by external stimuli (Yang et al. 2023; Gorain et al. 2020). Ongoing research in the field of PFC nanoemulsions is focused on optimizing their formulation, exploring novel stimuli-responsive behaviors, and developing strategies for targeted drug delivery to enhance the efficacy of therapies and reduce side effects. Low-boiling point perfluorocarbon nanodroplets, such as perfluorohexane (PFH) droplets, have been investigated for various biomedical applications, particularly in the field of ultrasound imaging and therapy. The low-boiling point characteristic of these PFH nanodroplets refers to their tendency to transition from a liquid to a gas phase at relatively low temperatures compared to other substances. This property allows the nanodroplets to undergo rapid phase transitions when exposed to certain stimuli, such as ultrasound. When PFH nanodroplets are exposed to ultrasound, they undergo a process called acoustic droplet vaporization (ADV). ADV is a phenomenon where the nanodroplets rapidly vaporize and transition into gas-filled microbubbles in response to ultrasound waves (Yang et al. 2023; Kwan and Borden 2012).

PFH nanodroplets stimulated by ultrasound hold promise for cancer drug delivery. These nanodroplets can be designed to accumulate at tumor sites and, when triggered by ultrasound, vaporize and release encapsulated drugs specifically within the tumor region. This localized drug release minimizes systemic exposure and reduces off-target effects. Ultrasound allows for precise control over the timing and extent of drug release, enhancing therapeutic efficacy and reducing toxicity. Moreover, the vaporization of PFH nanodroplets induces mechanical effects that enhance drug penetration into the tumor tissue, improving the effectiveness of chemotherapy. The combination of PFH nanodroplets and ultrasound can be synergistically combined with other therapies and is noninvasive, repeatable, and compatible with existing ultrasound technology. Ongoing research in this area aims to optimize these systems for improved cancer treatment outcomes (Samani 2023). While PFH nanodroplets have shown promise for drug delivery applications, there are some drawbacks associated with their use. Firstly, PFH nanodroplets have a limited payload capacity, restricting the amount of the drug that can be loaded into them. This can pose challenges for delivering high doses or large therapeutic molecules. Secondly, the circulation half-life of PFH nanodroplets is relatively short, as they can be quickly cleared from the bloodstream by the body's immune system or other clearance mechanisms. This may require repeated doses or alternative administration strategies to maintain therapeutic concentrations. Additionally, there is a potential risk of immunogenicity, as the immune system may recognize PFH or the encapsulation materials as foreign substances, leading to immune reactions or toxicity. Stability is another concern, as PFH nanodroplets can be prone to coalescence or premature vaporization, requiring careful formulation optimization. Lastly, PFH nanodroplets generally lack active targeting capabilities, necessitating the incorporation of additional targeting ligands or strategies to achieve specific targeting of cells or tissues (Abou-Saleh et al. 2016). Overcoming these drawbacks will require addressing payload capacity, circulation half-life, immunogenicity concerns, stability challenges, and enhancing targeting strategies to fully leverage PFH for drug delivery systems. In our previous study (Baghbani et al. 2017), chitosan was used as a shell to enhance nanodroplet stability and enhance drug bioavailability. Chitosan, being a cationic polymer, can form electrostatic interactions with negatively charged components, including perfluorohexane and curcumin. This interaction contributes to the stability of the nanodroplets, preventing premature drug release and facilitating controlled drug delivery. In another study (Baghbani et al. 2016), we investigated the properties and potential of doxorubicin-loaded alginate-stabilized perfluorohexane (PFH) nanodroplets as multifunctional drug delivery systems. The nanodroplets, synthesized using a nanoemulsion preparation method, exhibited ultrasound responsivity, imaging capabilities, and therapeutic effects. In the Gao et al. (2021) study, the chitosan/alginate complex was investigated as the stable shells of the PFH nanodroplets. The use of chitosan and alginate together provided pH-sensitive properties based on the acidic characteristics of tumor tissues. Chitosan and alginate formed a polyelectrolyte complex due to their opposite charges, enabling targeted and sustained drug release in response to the tumor microenvironment. Despite the advantages and suitability for certain applications of chitosan and alginate, chitosan and alginate showed moderate stability. This instability may affect the long-term storage and shelf life of the nanodroplets, making them more prone to degradation or changes in their structural integrity over time. It is important to consider this aspect when designing and formulating PFH nanodroplets with chitosan or alginate shells to ensure their efficacy and stability throughout their intended shelf life. Gelatin is a biocompatible, versatile, and readily available material that offers excellent stability and mechanical properties, making it a suitable choice for the shell structure of PFH nanodroplets. Its ability to form a stable shell around the PFH core provides protection against degradation and ensures the long-term stability of the nanodroplets. Furthermore, gelatin can be easily modified or functionalized to enhance specific properties such as drug loading capacity, controlled release, and targeting abilities. Additionally, gelatin is compatible with various fabrication techniques, allowing for the efficient production of uniform and well-defined nanodroplets. Its biodegradability and non-toxic nature further contribute to its suitability for biomedical applications. Overall, gelatin demonstrates great potential as a shell material in PFH nanodroplets, enabling their successful use in drug delivery and other therapeutic applications (Shende and Jain 2019).

In this study, ultrasound-responsive nanodroplets were developed as a type of emerging smart drug delivery system that enables nano-therapy for various diseases, especially cancer. Here, we developed multifunctional smart gelatin/perfluorohexane nanodroplets loaded with berberine chloride for targeted drug delivery. Berberine chloride is a naturally occurring alkaloid extracted from various plants, including Berberis vulgaris and Coptis chinensis. It has been traditionally used in Chinese and Ayurvedic medicine for its antibacterial, antifungal, and anti-inflammatory properties. In recent years, berberine chloride has gained increasing attention for its potential as an anti-cancer agent. Studies have shown that berberine chloride can inhibit cancer cell proliferation, induce apoptosis, and sensitize cancer cells to chemotherapy. Its ability to target multiple signaling pathways involved in cancer development makes it a promising candidate for cancer treatment. However, its clinical use has been limited by its poor solubility and low bioavailability (Majidzadeh, et al. 2020). The encapsulation of berberine chloride within gelatin/perfluorohexane nanodroplets improves its solubility and stability, allowing for better drug dispersion and preventing precipitation. This enhances the bioavailability of berberine chloride and improves its therapeutic efficacy. Gelatin as the shell material provides a protective barrier around the berberine chloride, shielding it from degradation and enzymatic degradation in the biological system. This protection ensures the sustained release of berberine chloride and prolongs its presence at the target site, optimizing its therapeutic effect. Furthermore, the small size and surface properties of gelatin/perfluorohexane nanodroplets facilitate their accumulation at the target site through the enhanced permeability and retention (EPR) effect. This passive targeting mechanism allows for the preferential accumulation of the nanodroplets in tumor tissues, where berberine chloride can exert its therapeutic activity.

## Methods

### Materials

Gelatin (type B, Bloom250), Tween 80 as a surfactant, glutaraldehyde (25% (v/v) in water), and sodium hydroxide were purchased from Merck (Darmstadt, Germany). Berberine chloride and perfluorohexane were supplied by Sigma-Aldrich (St. Louis, MO, Canada). Deionized water was used for all the experiments. All other chemicals were of reagent grade and obtained from Sigma-Aldrich.

### Preparation of PFH nanodroplets with a gelatin shell containing berberine chloride

The nanoemulsion method was used for the synthesis of nanodroplets in this study. We have revised our previous methodology to suit our new study (Baghbani et al. 2016). First, gelatin solutions were prepared in different concentrations (Table 1) in deionized water under stirring on a magnetic stirrer (IKA RCT basic) at 500 rpm and at 50°C for 10 min to obtain a completely clear solution. NaOH (0.1 M) was added to adjust the pH to 6, which prevented deposition of gelatin. After cooling the gelatin to room temperature, the polymeric solutions were filtered. Berberine chloride was dissolved in deionized water with a magnetic stirrer. Surfactant solution (Tween 80 0.5%v/v)

**Table 1** Formulations and process variables on the production of nanodroplets loaded with berberine chloride using PFH and gelatin

| Formulation code | Gelatin (%w/v) | Berberine chloride (mg) | PFH (μL) | Tween 80 (%v/v) | Homogenization speed (rpm) |
|---|---|---|---|---|---|
| *Variable: polymer concentration (%w/v)* | | | | | |
| $A_a$ | 2.3 | 5 | 150 | 0.5 | 24,000 |
| $A_b$ | 2 | 5 | 150 | 0.5 | 24,000 |
| $A_c$ | 1.67 | 5 | 150 | 0.5 | 24,000 |
| *Variable: amount of drug (mg)* | | | | | |
| $B_a$ | 1.67 | 4 | 150 | 0.5 | 24,000 |
| $B_b$ | 1.67 | 5 | 150 | 0.5 | 24,000 |
| $B_c$ | 1.67 | 7 | 150 | 0.5 | 24,000 |
| *Variable: amount of perfluorohexane (μl)* | | | | | |
| $C_a$ | 1.67 | 5 | 150 | 0.5 | 24,000 |
| $C_b$ | 1.67 | 5 | 200 | 0.5 | 24,000 |
| *Variable: Surfactant content (% v/v)* | | | | | |
| $D_a$ | 1.67 | 5 | 150 | 0.3 | 24,000 |
| $D_b$ | 1.67 | 5 | 150 | 0.5 | 24,000 |
| *Variable: Homogenizer speed (rpm)* | | | | | |
| $F_a$ | 1.67 | 5 | 150 | 0.5 | 17,000 |
| $F_b$ | 1.67 | 5 | 150 | 0.5 | 24,000 |

All components of the above formulations were added to 3 ml of deionized water and drug solution were added to perfluorohexane and they were homogenized for 2 min at 24,000 rpm using Ultra Turrax T25 homogenizer. Then, the prepared polymer solution was added dropwise to the above emulsion under homogenization at 13,000 rpm for 5 min. Glutaraldehyde solution was then added to the emulsion as the cross-linking agent dropwise and stirred at 5000 rpm. The cross-linking process was terminated by the addition of 8 mL of aqueous sodium metabisulfite solution (1.6% w/v). In order to prepare nanodroplets with suitable size in the nano range and high drug loading efficiency, various formulations were tested by adjusting different process parameters. The size of the nanodroplets, drug loading efficiency, and drug release kinetics all depend on these parameters. In order to optimize these properties, several factors were considered such as polymer concentration, drug content, surfactant concentration, and perfluorohexane amount. Additionally, process parameters like homogenizer rate were also investigated. Table 1 summarizes different combinations in the synthesis procedure.

## Characterization of berberine chloride-loaded nanodroplets

### Particle size, size distribution and zeta potential

The average particle size, polydispersity index (PDI), and zeta potential of the nanodroplets were measured via dynamic light scattering (DLS) using a Zetasizer Nano ZS (VASCO2, Cordouan Tech, France). The nanodroplet emulsions were diluted 1:4 with deionized water before the measurement.

### Morphology of drug-loaded nanodroplets

Transmission electron microscopy (TEM) is a powerful technique used for imaging and characterizing nanoscale materials and structures. While it is commonly used for solid samples, it can also be used to study nanodroplets. The morphology and structure of the nanodroplets were observed using a TEM (FEI TEC9G20, Hitachi, Tokyo, Japan). To prepare the samples, a drop of the emulsion was poured on a 400-mesh carbon-coated copper grid and dried at room temperature. To enhance contrast for imaging, the sample was treated with 1% alkaline phosphotungstic acid (PTA) for several minutes and allowed to dry at room temperature before analysis.

### Fourier transform infrared (FTIR) spectroscopy

FTIR spectroscopy provides information about the chemical composition of the drug delivery system. It can identify functional groups present in the system, allowing it to confirm the presence of specific compounds, polymers, or drug molecules. Nanodroplets were lyophilized for preparing micronized KBr pellets. The samples were scanned in the 4500–400 $cm^{-1}$ spectral range.

### Evaluation of encapsulation efficiency

To determine the encapsulation efficiency (EE), drug-loaded nanodroplet emulsions were centrifuged at 15,000 rpm for 40 min and 4 °C (Universal320R, Hettich) to do the differential separation of the encapsulated drug from the non-encapsulated drug. Carefully collect the supernatant, which contains the non-encapsulated drug, without disturbing the pellet of nanodroplets. The amount of non-encapsulated berberine chloride was quantified by measuring its absorbance at 418 nm using a UV–Vis spectrophotometer (UV-5100PC Shanghai Metash Instruments Co., Ltd., China) and comparing the results to a standard curve prepared beforehand. Drug entrapment efficiency (EE) was determined by calculating the ratio of the amount of non-encapsulated drug concentration to the total amount of drug used for nanodroplet preparation, using the following equation:

$$\text{Entrapment efficiency (\%)} = \frac{(M_0) - (M_f)}{M_0} \times 100$$

$M_0$ is the total amount of drug added to the formulation and $M_f$ is the amount of free drug in the supernatant calculated based on the constructed calibration curve using standard solutions of known drug concentrations.

### Stability of nanodroplets

The experiment involved suspending nanodroplets in PBS with a pH of 7.4 and incubating them at 4 °C for specific time intervals. The stability of the nanodroplets was determined by measuring any changes in their size and drug entrapment efficiency over a month.

### In vitro evaluation of drug release

*Evaluation of drug release under passive conditions* In this study, we employed the dialysis method to investigate the kinetics of drug release from nanodroplets. The release behavior of berberine chloride was evaluated at two different pH values (7.4 and 5.5) to simulate the physiological conditions and the slightly acidic extracellular matrix of most tumors. The pH range in tumor tissues can vary, but it is generally lower (more acidic) compared to healthy tissues. The pH range in tumors is typically reported to be between 6.5 and 7.0, with some studies suggesting even lower pH values. The extracellular pH in the tumor microenvironment tends to be more acidic, while intracellular pH can be closer to neutral or slightly alkaline in some tumor cells (Zhang et al. 2010).

To conduct the experiment, 2 ml of the optimized nanodroplet emulsion containing berberine chloride was placed in a dialysis bag (molecular weight cutoff 12,000 Da), and the bags were then immersed in 10 ml of PBS with pH 7.4 and citrate buffer with pH 5.5. A magnetic stirrer was used to maintain conditions at a constant temperature of 37 °C and a speed of 100 rpm. At specific time intervals (0–8 and 24 h), 2 ml of the buffer was removed and replaced with an equal volume of fresh buffer. The amount of berberine chloride released was determined by a UV–Vis spectrophotometer (UV-5100PC Shanghai Metash Instruments Co., Ltd., China) at 418 nm.

In order to determine the accumulated release of berberine chloride from nanodroplets, the following formula was used:

$$\text{Accumulated release (\%)} = \frac{\left[V \sum n - i(C_i + V_0 C_n)\right]}{m_{\text{drug}}}$$

where $V$ is the sampling volume, $V_0$ is the initial volume, $C_i$ and $C_n$ are the berberine chloride concentrations, $i$ and $n$ are the sampling times, and $m_{\text{drug}}$ is the mass of berberine chloride in nanodroplets.

*Evaluation of stimulated drug release* To investigate the effects of ultrasound waves on the release of berberine chloride from nanodroplets, a nanoemulsion of nanodroplets was applied onto a culture dish as a scaffold and exposed to a low-frequency ultrasound converter (SM3678B, Shrewsbury Medical Ltd., Shropshire, UK),150 kHz, 0.2 W/cm$^2$, continuous mode, and 0.5 cm distance from the dish with a cross-sectional area of 5 cm$^2$ probes inside the chamber and high-frequency of ultrasound converter (Enraf Nonius, Rotterdam, The Netherlands), 1 MHz, 0.2 W/cm$^2$, continuous mode, and 0.5 cm distance from the dish ultrasound waves for varying durations of 30 s, 5 min, and 10 min. Previous studies have shown that the frequencies of 150 kHz and 1 MHz are safe for cells and do not cause cell lysis (Barati and Mokhtari-Dizaji 2010). The effective radiation area (ERA) was 30 mm$^2$ with a geometric cross-sectional area of 5 cm$^2$ probes. Following exposure, the nanodroplet solution was centrifuged at 15,000 rpm, and the concentration of the released drug in the supernatant was measured as described above. The drug release obtained from both low and high frequencies was compared.

### Statistical analysis
All experiments were performed at least in triplicate. Two-way ANOVA and Tukey's multiple comparison tests were used for the statistical analysis with significant reports when $P < 0.05$.

## Results
The effect of different processes and structural parameters on the average particle size, zeta potential, entrapment efficiency, and cumulative drug released from different formulations of nanodroplets are collected in Table 2. In what follows, the results were discussed in detail.

## Discussion
### Effect of polymer concentration variation
Changes in particle size and drug loading were observed by altering the polymer concentration of the emulsion system in formulations Aa, Ab, and Ac. With a decrease in the weight percentage of gelatin, a reduction in particle size from 635.1, 458, to 281.7 nm was achieved, as shown in Table 2. Also, as expected, with decrease in polymer concentrations (2.3, 2, and 1.67% w/v), less drug was trapped between polymer chains, and drug loading decreased from 73.3 to 69.2% and 66.8%. The polymer concentration in the emulsion system has a significant impact on both particle size and drug loading (Lagreca, et al. 2020). Lower polymer concentrations result in smaller particle sizes, leading to a reduction in drug loading. In emulsion systems, the polymer acts as a stabilizer, helping to maintain the dispersion of the drug particles within the system. When the polymer concentration is high (as in formulation Aa), there is more polymer available to stabilize the particles, resulting in larger particle sizes (Krishnamachari et al. 2007). Since polymer chains can act as a matrix to trap the drug molecules, increasing the polymer concentration (as in formulation Aa) in the emulsion system provides more available polymer chains, which enhances the capacity for drug encapsulation. This can prevent or delay the saturation of drug encapsulation, leading to higher drug loading and improved encapsulation efficiency. On the other hand, the amount of berberine chloride released from formulations Aa, Ab, and Ac during 24 h was 10.2%, 12.1%, and 15.4%, respectively. The higher percentage of berberine chloride released from formulation Ac (with smaller droplets) compared to formulation Aa (with larger droplets) suggests a slower percent drug release from formulation Ac. The formation of larger droplets can have implications for percent drug release. Typically, smaller droplets have a larger surface area-to-volume ratio, allowing for faster drug release. When the amount of polymer increases and the viscosity of the emulsion system increases, it can limit the diffusion of drug molecules from the organic phase to the aqueous phase. The increased viscosity and larger particle size, as a result of higher polymer concentration, create a longer diffusion path for drug molecules to travel

**Table 2** The effect of different formulations and structural parameters on the average particle size, polydispersity index (PDI), zeta potential, encapsulation efficiency (EE), and cumulative drug release

| Formulation code | Variable | Mean particle size (nm) | Mean PDI | Zeta-potential | EE (% ± SD) | Cumulative drug released within 24 h (% ± SD) |
|---|---|---|---|---|---|---|
| | *Gelatin concentration (%w/v)* | *Variable: polymer concentration (%w/v)* | | | | |
| $A_a$ | 2.3 | 635.1 | 0.30 | 30.61 | 73.3 ± 2.3 | 10.2 ± 0.3 |
| $A_b$ | 2 | 458 | 0.26 | 31.50 | 69.2 ± 2.1 | 12.1 ± 0.4 |
| $A_c$ | 1.67 | 281.7 | 0.24 | 31.52 | 66.8 ± 1.7 | 15.4 ± 0.2 |
| | *Berberine chloride (mg)* | *Variable: drug amount (mg)* | | | | |
| $B_a$ | 4 | 278.2 | 0.32 | 31.63 | 63.4 ± 1.5 | 12.4 ± 0.1 |
| $B_b$ | 5 | 281.7 | 0.24 | 31.52 | 66.8 ± 1.7 | 15.4 ± 0.2 |
| $B_c$ | 7 | 351.2 | 0.19 | 31.45 | 53.7 ± 1.9 | 14.1 ± 0.4 |
| | *PFH (µL)* | *Variable: amount of perfluorohexane (µl)* | | | | |
| $C_a$ | 150 | 281.7 | 0.24 | 31.52 | 66.8 ± 1.7 | 15.4 ± 0.2 |
| $C_b$ | 200 | 363.3 | 0.4 | 31.38 | 71.8 ± 3.5 | 12.9 ± 0.3 |
| | *Tween 80 (%v/v)* | *Variable: surfactant content (% v/v)* | | | | |
| $D_a$ | 0.3 | 366.4 | 0.38 | 31.35 | 59.9 ± 3.8 | 10.9 ± 0.3 |
| $D_b$ | 0.5 | 281.7 | 0.24 | 31.52 | 66.8 ± 1.7 | 15.4 ± 0.2 |
| | *Homogenization speed* | *Variable: Homogenizer speed (rpm)* | | | | |
| $F_a$ | 1700 | 467.3 | 0.22 | 31.32 | 45.3 ± 3.5 | 11.2 ± 0.3 |
| $F_b$ | 2400 | 281.7 | 0.24 | 31.52 | 66.8 ± 1.7 | 15.4 ± 0.2 |

All components of the above formulations were added to 3 ml of deionized water from the particle to the surrounding aqueous phase (Görner, et al. 1999). This longer path can impede the drug's ability to diffuse out of the particles and into the aqueous medium. As a consequence, the drug molecules may have a higher tendency to be trapped within the particles, leading to increased drug retention or reduced drug release. This can result in a higher amount of drug being trapped in the particles, as observed in the formulations with larger particle sizes due to increased polymer concentration. According to the nanodroplet size, EE, and drug release profile, Ac formulation with 1.67% w/v gelatin was selected as an optimum polymer concentration. In addition, PDI in all samples has a value between 0.2–0.3. PDI is a parameter used to characterize the size distribution of particles, including nanodroplets. A PDI value between 0.1 and 0.3 is often considered reasonable for many applications, including drug delivery systems (Kushwaha et al. 2019; Focarete and Tampieri 2018).

The positive zeta potential value of nanodroplets is reported in Table 2. Gelatin has an amphoteric nature, meaning it can have both positive and negative charges depending on the pH of the surrounding medium. At its isoelectric point (pI), gelatin has a near-neutral zeta potential. However, in most practical drug delivery applications, the pH of the medium is not at the pI of gelatin, and therefore, gelatin nanodroplets tend to have a slightly positive zeta potential in aqueous solutions. The zeta potential measurements were in the range of 30–32 and changes in formulations did not influence them.

### Effect of drug amount variation

The particle sizes of the core–shell nanocarriers increased with higher drug loading concentrations. The sizes recorded for drug loadings of 4 mg, 5 mg, and 7 mg were 278.2 nm, 281.7 nm, and 351.2 nm, respectively. The observed increase in particle size indicates that higher drug loading leads to larger nanocarrier structures. Interestingly, the entrapment efficiencies showed an unexpected trend. The entrapment efficiencies for the drug loading concentrations of 4 mg, 5 mg, and 7 mg were 63.4 ± 1.5%, 66.8 ± 1.7%, and 53.7 ± 1.9%, respectively. Contrary to the particle size results, the entrapment efficiency decreased with higher drug loading. This unexpected finding suggests that factors other than drug loading concentration may be influencing the encapsulation process. The increase in particle size with higher drug loading can be attributed to the increased amount of drug incorporated into the nanocarrier system. While the decrease in entrapment efficiency with higher drug concentration is intriguing and warrants further investigation. It is possible that at higher drug concentrations, the drug molecules might have a tendency to aggregate

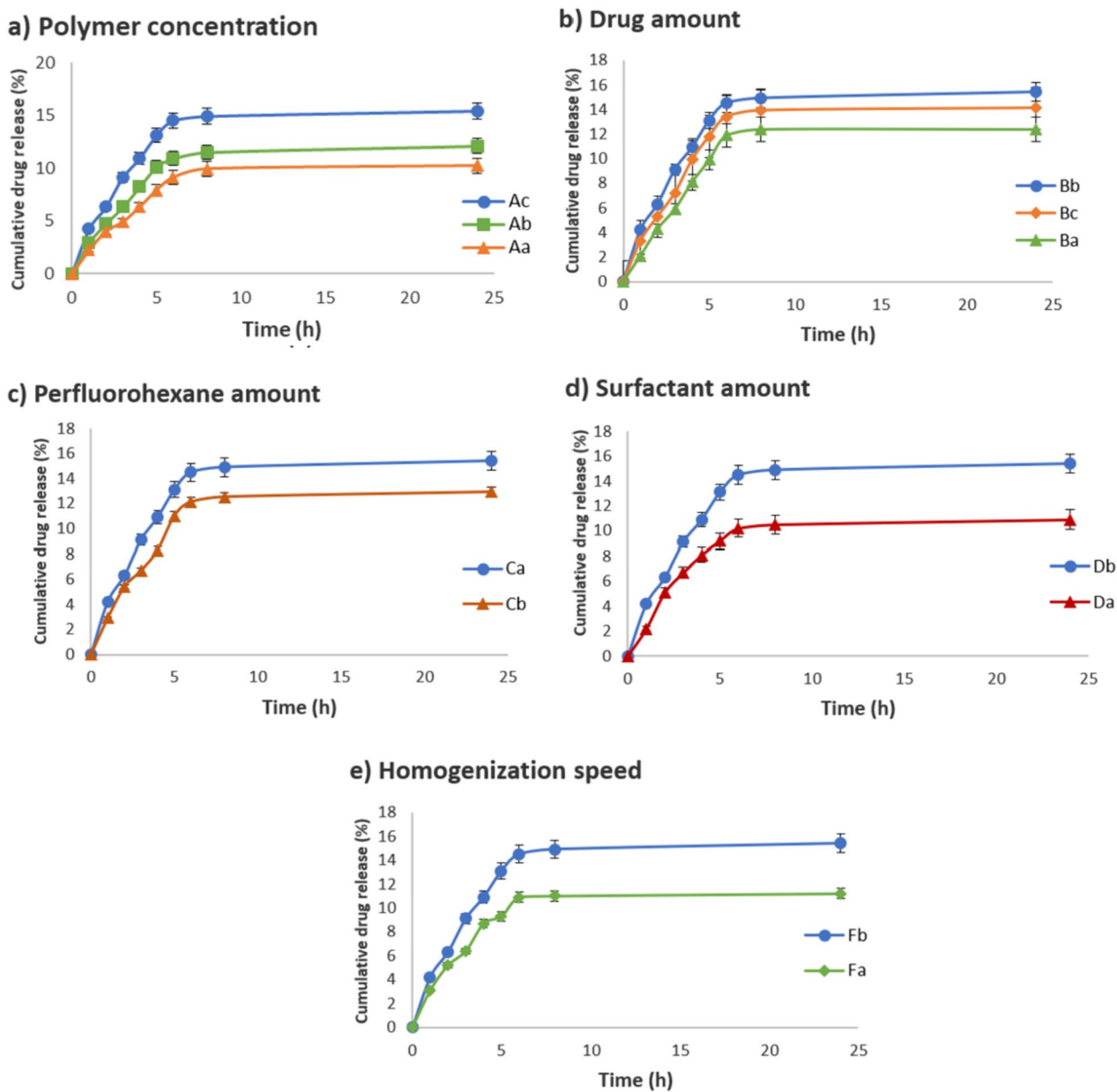

**Fig.1** Cumulative drug-released percentage within 24 h in different formulations. Variables included **a** polymer concentration, **b** drug amount, **c** perfluorohexane amount, **d** surfactant amount, and **e** homogenization rate. Polymer amount, perfluorohexane amount, surfactant amount, and homogenization rate showed significant effects on drug-released profiles

and precipitate on the shell surface so reducing their incorporation into the core–shell nanocarriers (Focarete and Tampieri 2018). As expected, the formulation with higher entrapment efficiency (Ba formulation) demonstrated a higher cumulative drug release within the initial 24-h period (Fig. 1b). This finding suggests a correlation between entrapment efficiency and drug release kinetics in the studied system. This can be attributed to the improved retention of the drug within the nanocarrier, resulting in a more efficient drug release process.

### Effect of perfluorohexane amount variation

Perfluorohexane plays a crucial role in core–shell nanocarriers as the core material. It enables the generation of microbubbles upon ultrasound stimulation, providing contrast enhancement and targeted therapy. With its low surface tension, high volatility, biocompatibility, and chemical stability, perfluorohexane facilitates efficient dispersion, controlled release, and gas exchange, making it a versatile and valuable component in nanocarrier design (Yazdian Kashani et al. 2020). The investigation in this study focused on the effect of increasing the amount of perfluorohexane, an important component of the system's core, on the size of nanodroplets and drug loading. The findings revealed that an increase in the amount of perfluorohexane resulted in an increase in particle size, as shown in Table 2. Nanodroplets obtained from the Ca and Cb formulations had mean sizes of 281.7 nm and 363.3 nm, respectively. This increase in particle size can be attributed to the emulsions containing a higher oil phase content, in this case, perfluorohexane. When emulsions have a higher oil phase content, there may not be enough surfactant available to adequately cover the surface of each droplet and prevent them from merging. Consequently, larger droplets are formed, resulting in a lower specific surface area and larger particle sizes (Fuhrmann et al. 2019). It appears that alterations in the perfluorohexane content do not greatly impact drug loading. This suggests that the majority of drugs are trapped effectively within the gelatin shell. This finding suggests that the drug encapsulation process is primarily determined by the properties of the gelatin shell rather than the amount of perfluorohexane in the core.

With the increase in the oil phase, the percent drug release decreased from 15.4 to 12.9% in Ca and Cb, respectively (Fig. 1c). The reduction in specific surface area as a result of increasing the size of the nanocarrier can affect the drug release kinetics. A lower surface area may result in decreased drug diffusion and slower drug release from the nanodroplets, leading to a decreased percentage of drug release (Yasmin et al. 2017). These findings highlight the importance of considering the composition and formulation parameters in drug delivery systems. The amount of perfluorohexane plays a crucial role not only in particle size but also in drug release kinetics. Optimizing the formulation parameters can help achieve desired drug release profiles and enhance the therapeutic efficacy of the delivery system. Increasing the content of perfluorohexane in the system creates nanodroplets with larger cores and thinner shells. The thinner the shell and the larger the particle size result in lower loading of the drug. In addition, smaller nanodroplets need less energy to become microbubbles so the heat generation after ultrasound stimulation is reduced (Gao et al. 2008).

**Effect of tween 80 concentration**

The selection and concentration of the surfactant can influence the particle size of the nanocarriers. The surfactant molecules can act as steric or electrostatic barriers, preventing the aggregation or coalescence of nanodroplets and leading to smaller particle sizes. Additionally, surfactants can influence the emulsion droplet size, which in turn affects the final particle size of the nanocarriers (Wanx and Lee 1974; Quintanar-Guerrero et al. 1996). Tween 80 or polysorbate 80 is a nonionic surfactant that has excellent emulsifying properties, making it effective in stabilizing oil-in-water emulsions. It can reduce the interfacial tension between oil and water phases, leading to the formation of stable emulsions with small droplet sizes. This property is particularly advantageous in the preparation of nanocarriers, as it helps in achieving uniform and stable nanodroplet dispersions. In addition, its lipophilic portion can improve the solubility and dispersibility of poor water solubility drugs like berberine chloride in aqueous media. By forming micelles or solubilizing the drug, Tween 80 can enhance the drug's bioavailability and facilitate its encapsulation within the nanocarrier system (Kwon et al. 2020). To explore how varying surfactant concentrations affect particle size, drug loading rate, and percent drug release, two distinct concentrations of the surfactant were employed. By adjusting the concentration of surfactant in the formulation, the particle size and encapsulation efficiency can be modified. Increasing the surfactant concentration (0.3 and 0.5% volume of tween 80) resulted in smaller particle sizes in Da (366.4 nm) and Db (281.7 nm) formulations, and increased encapsulation efficiency in formulations Da (59.9%) and Db (66.8%). The higher percentage of Tween 80 in the formulation can promote the formation of smaller and more uniform nanodroplets. As a surfactant, Tween 80 reduces the interfacial tension between the oil and water phases, allowing for better dispersion and stabilization of the hydrophobic compound within the aqueous medium. This leads to the formation of smaller droplets or nanodroplets during emulsification or self-assembly processes. The presence of Tween 80 at higher concentrations also enhances the encapsulation efficiency of berberine chloride (Table 2). The surfactant molecules can interact with the hydrophobic compound, forming micelles or complex structures that facilitate its entrapment within the nanocarrier system. This improves the loading capacity and reduces the potential loss of the drug during the formulation process. The amount of berberine chloride released from the system during 24 h of the mentioned formulations were 10.9% and 15.4%, respectively (Fig. 1d). It was anticipated that the drug release would be greater in the Db formulation due to the smaller size of the nanocarriers, as previously mentioned in "Effect of perfluorohexane amount variation" section.

**Effect of speed of homogenizer**

The main role of homogenizer is to facilitate the emulsification and dispersion of the components to achieve a stable and uniform formulation. The homogenizer applies mechanical shear forces to break down the immiscible component (perfluorohexane (oil phase)) and aqueous components (berberine chloride, Tween 80, gelatin). It helps in overcoming the interfacial tension between these components and creates smaller droplets of the oil phase dispersed in the aqueous phase. In the synthesis of nanodroplets, varying homogenization rates of 17,000 rpm and 24,000 rpm were utilized. It was noted that the particle size decreased as the velocity increased, which was in line with expectations (Table 2). The high-speed rotation of the homogenizer generates intense shear forces, resulting in the fragmentation of larger droplets into smaller droplets. Smaller droplets provide a larger interfacial area, which is beneficial for encapsulating the hydrophobic berberine chloride within the perfluorohexane droplets and improving the stability of the formulation (Budhian et al. 2007). Furthermore, when the homogenization speed was increased to 24,000 the encapsulation efficiency also increased to 66.8%. By increasing the homogenizer speed, these factors collectively contribute to improved emulsification, enhanced mixing, and optimized surface coverage, resulting in increased encapsulation efficiency of the drug within the nanodroplets. Smaller droplets provide more opportunities for the hydrophobic berberine chloride molecules to partition and become encapsulated within the oil phase (perfluorohexane) of the nanodroplets, leading to improved encapsulation efficiency. More efficient mixing ensures better distribution of the hydrophobic drug within the oil phase and improves the chances of drug molecules being encapsulated rather than remaining in the aqueous phase. In addition, the increased homogenizer speed promotes better coverage of the droplet surfaces by the surfactant (Tween 80) molecules. A higher speed allows for more effective adsorption and arrangement of surfactant molecules at the oil–water interface, forming a stable and well-structured interfacial layer. This improved surface coverage helps to stabilize the nanodroplets, prevent coalescence, and enhance the encapsulation of the hydrophobic drug (Zamani et al. 2022). The nanodroplets synthesized at 24,000 rpm had a higher percent drug release (15.4%) compared to the larger nanodroplets obtained at 17,000 rpm (11.2%) after 24 h (as shown in Fig. 4). It was anticipated that the drug release would be greater in the Fb formulation due to the smaller size of the nanocarriers, as previously mentioned in "Effect of perfluorohexane amount variation" section.

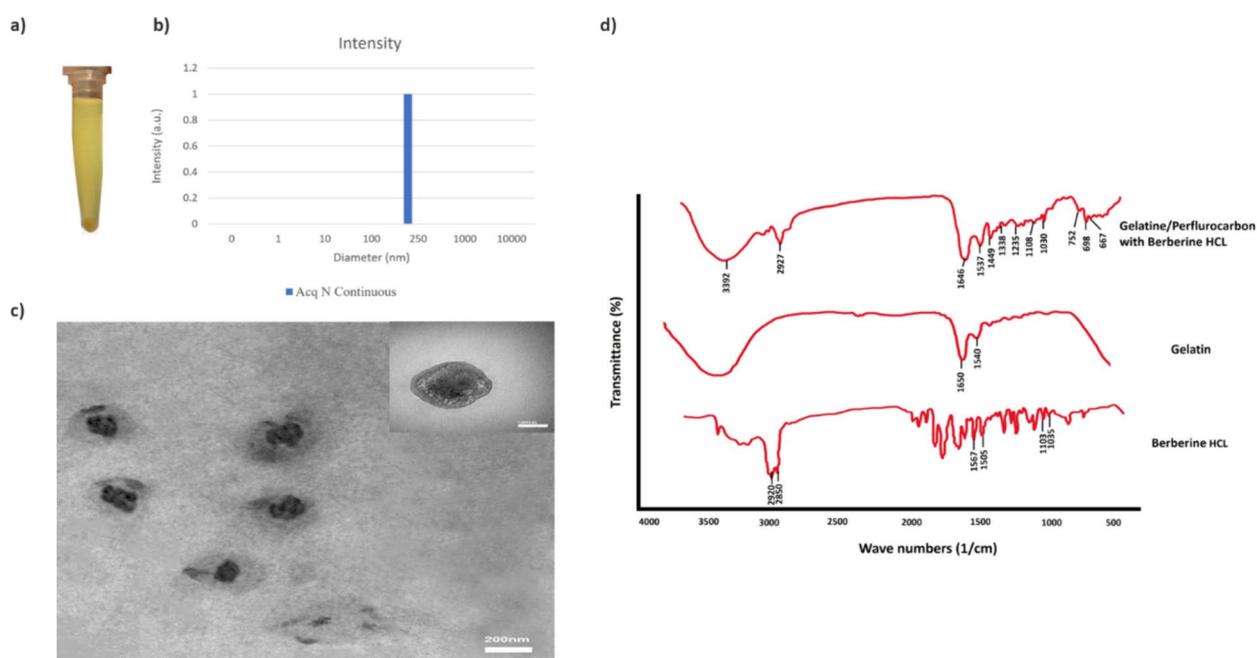

**Fig. 2** Physicochemical characterization of optimum formulation (Ac). **a** Precipitated berberine chloride-loaded nanodroplets after centrifugation that separated for further analyses. **b** The size distribution of nanodroplets obtained from DLS. The graph showed the particle size and high monodispersity of the nanodroplets. **c** TEM images of gelatin nanodroplets containing berberine chloride. Nanodroplets had a distinct core material (PHF) surrounded by a gelatin shell. High-magnitude detail inset: Inset in the top-left corner showcases a high-magnitude view of one nanodroplet within the TEM image. **d** The FTIR spectrum of Ac formulation besides the drug and gelatin

**Physicochemical properties of optimum formulation**

Among various formulations tested for synthesizing nanodroplets containing berberine chloride with a gelatinous shell, the Ac formulation with an average particle size of 281.7 nm and an encapsulation efficiency of 66.8% is the optimal formulation for continuing further testing. The size and morphology of drug-loaded nanodroplets were observed using DLS and TEM (Fig. 2b, c). The TEM image displays nanodroplets with a semi-circle shape and a size of approximately 280 nm. This size corresponds with the DLS results. In addition, DLS showed an acceptable polydispersity index. Upon observation using TEM, it was evident that the nanodroplets had a distinct core material (PHF) surrounded by a gelatin shell. The presence of a suitable amount of perfluorohexane in the system can create nanodroplets whose core–shell structure can be easily identified in the TEM image.

FTIR analysis of the nanocarrier loaded with berberine chloride can provide valuable information regarding the presence of gelatin and berberine chloride, chemical interactions between them, structural changes, and the stability of the composite material. The FTIR spectra of gelatin are characterized by the main amide bands. The amide I band in the frequency range of 1650–1630 $cm^{-1}$ represents the C=O stretching vibrations of the amide bonds in the protein backbone. The amide II in the range of frequency of 1560–1530 $cm^{-1}$ corresponds to N–H bending and C–N stretching vibrations (indicated in Fig. 2d). It provides information about the secondary structure of gelatin, including α-helices, β-sheets, and random coils (Hermida-Merino 2022). The presence of these peaks in the nanodroplet formulation with small shifts confirmed that there are not any changes in the secondary structure of the gelatin shell caused by the presence of berberine chloride. In addition, a wide peak at 3300–3500 $cm^{-1}$ region may be attributed to N–H stretching vibrations of the amide group of the gelatin and is a characteristic Amide A band (Khan et al. 2016) Berberine HCl exhibited significant peaks at 2920–2930 $cm^{-1}$ that represents stretching vibrations of the aliphatic carbon–hydrogen bonds in the molecule. This peak presented at 2927 $cm^{-1}$ in the nanodroplet formulation. The peak at 1537 $cm^{-1}$ corresponds to the stretching vibrations of the aromatic C=C rings and the peak at 1030 $cm^{-1}$ indicates the C–H vibrations present in berberine chloride (Lam et al. 2012). The spectrum of nanodroplets loaded with berberine chloride showed the characteristic peaks of berberine HCl and gelatin.

**Passive drug release**

It is generally desirable to optimize the design of stimuli-responsive nanocarriers to achieve a high degree of control over drug release while minimizing passive release. The idea is to minimize the passive or uncontrolled release of the drug when it is not needed, thereby reducing potential side effects and improving the therapeutic efficacy. The optimum level of passive drug release will vary depending on the specific requirements of the application. In cancer treatments, it is crucial to minimize passive release to avoid unnecessary drug exposure to healthy tissues and reduce systemic toxicity (Adepu and Ramakrishna 2021). In Ac formulation, the cumulative release after 24 h was 15.4% at neutral pH. The level of release is slightly higher than our previous drug delivery system, which utilized alginate as a shell (Baghbani et al. 2016). Considering the structural properties of alginate and its ability to form a gel-like structure, it was expected that gelatin forms a denser and less porous matrix compared to alginate. This could potentially restrict the passive diffusion of drugs from the nanocarrier. The variation can depend on the specific formulation, preparation method, and other factors that influence the release kinetics of drugs from nanocarriers and lead to different outcomes than what might be expected based on general properties. It is also worth considering that gelatin is a protein-based material, and proteins can undergo changes in conformation or degradation over time, potentially altering the permeability of the gelatin shell and affecting drug release. It is important to note that the choice of nanocarrier material depends on various factors, including the specific drug being delivered, the desired release profile, and the target application. While protein-based materials like gelatin may have limitations in terms of passive release control, they can still be advantageous in certain contexts, such as for specific drug types or when a higher percent drug release is desired. The Ac formulations possessed significantly higher berberine chloride release at acidic pH (5.5, the pH of endosomal/lysosomal environment) compared to normal pH (7.4) (Fig. 3). The cumulative amount of drug release at acidic was about 27.2% which was significantly ($p < 0.05$) higher than normal pH (7.4). This behavior can be attributed to the pH-responsive characteristics of the formulations. Acidic pH can cause protonation of carboxyl groups in gelatin or other acidic functional groups. This protonation leads to the conversion of –COO– ionic groups to COOH groups. In other words, the carboxyl groups, which typically carry a negative charge at neutral pH, become neutralized due to the presence of excess protons in the acidic environment (Mahdian 2023). In an acidic environment, such as the endosomal/lysosomal compartments where the nanocarriers may be taken up by cells, the pH-sensitive components of the formulations may undergo changes, leading to increased drug release.

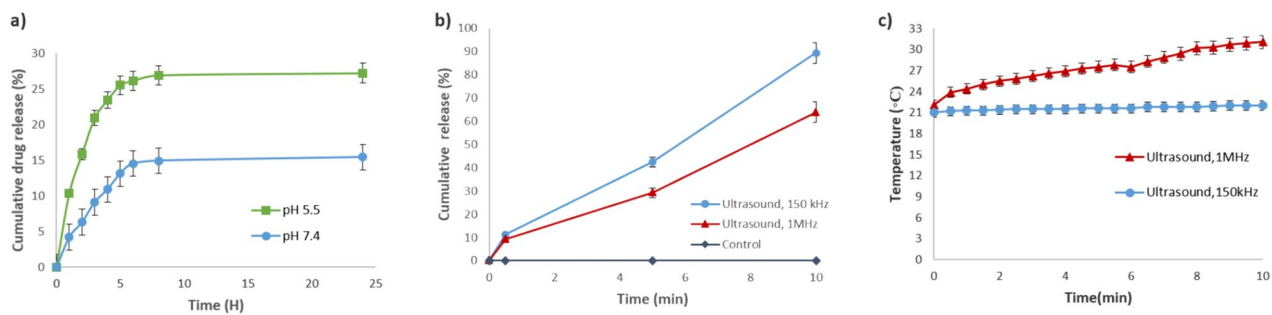

**Fig. 3** The comparison between passive and stimulated drug release of the optimum formulation and heat generation after ultrasound stimulation. **a** Berberine chloride release profile from $A_C$ formulation at neutral pH (7.4) and acidic pH (5.5). The acidic condition changed the release profile **b** The active release profile of berberine chloride from the optimal formulation (Ac) under low (150 kHz) and high (1MHz) frequency ultrasound for 10 min. **c** The temperature changing in the Ac formulation caused by the radiation of 1 MHz and 150 kHz ultrasound frequencies within 10 min, measured in degrees Celsius

**Ultrasound-induced drug release**

When perfluorocarbon nanodroplets are initially injected into the body, they remain in a liquid state. However, when exposed to ultrasound at a specific frequency and intensity, the nanodroplets can be triggered to vaporize and form gas bubbles. This vaporization process occurs due to the absorption of acoustic energy by the nanodroplets, leading to their rapid expansion and conversion into gas bubbles. The ADV process can induce the rapid expansion and disruption of the nanodroplets, facilitating the release of encapsulated drugs or therapeutic agents (Wang et al. 2012). The vaporization of perfluorohexane nanodroplets can generate mechanical forces that cause the disruption or fragmentation of the gelatin shell. The combination of ADV and gelatin shell disruption can lead to enhanced drug delivery. The vaporization process can create transient pores or openings in the gelatin shell, facilitating the controlled release and diffusion of berberine chloride. This can potentially improve the drug's bioavailability, increase its concentration at the target site, and enhance its therapeutic effects. The frequency of ultrasound plays a critical role in determining the threshold at which vaporization occurs. Different perfluorocarbon nanodroplets have specific vaporization thresholds that correspond to specific ultrasound frequencies. This threshold efficiency is typically in the range of $10^1$ to $10^3$ kHz (Zhao 2023). We studied the ADV effect for Ac formulation via injection through a thin needle and 150 kHz and 1 MHz ultrasound exposure at room temperature, in PBS solution for 10 min. Figure 3b clearly displays that the release of berberine chloride was significantly higher at 150kHz (89.4%) after 10 min compared to 1 MHz stimulation (63.9%). In general, lower ultrasound frequencies have a lower vaporization threshold for ADV compared to higher frequencies. This causes a higher percentage of nanodroplets to undergo vaporization at this frequency (Xu et al. 2018). Additionally, 150 kHz ultrasound frequency generates larger vaporized bubbles during ADV. These larger bubbles undergo expansion and contraction cycles without immediate collapse. Stable cavitation can enhance drug release and improve the efficiency of ADV while minimizing potential bioeffects. Lower ultrasound frequencies are more likely to promote stable cavitation. ADV, the process of perfluorocarbon nanodroplets vaporizing into gas bubbles, can generate localized heat. This heat is a result of the conversion of acoustic energy into kinetic energy, causing rapid expansion and subsequent collapse of the vaporized bubbles. The energy dissipation during bubble collapse can lead to the generation of heat in the surrounding medium. At high frequencies, the bubbles that form are larger in size and fewer in number. These bubbles are visible to the naked eye and are the main cause of heat generation (Husseini et al. 2000). Ultrasound has also proven to be safe and effective for therapeutic applications in clinical settings. In drug delivery systems stimulated by ultrasound, safety considerations like local heat generation and safe intensity are of paramount importance. In many medical applications, it is recommended to limit the increase in local tissue temperature to no more than 2–4 °C above the baseline body temperature. This range is generally considered safe to avoid damaging or injuring the surrounding tissues (Miller, et al. 2012). To explore the effects of ultrasound radiations, the temperature rise caused by radiations was monitored. The findings of the research confirmed our assumption, as there was a clear temperature increase in the 1 MHz radiation (as seen in Fig. 3c). However, it should be noted that this temperature increase can be harmful to the surrounding tissues and should be avoided. As a result, 150 kHz radiation is safe and effective for drug release.

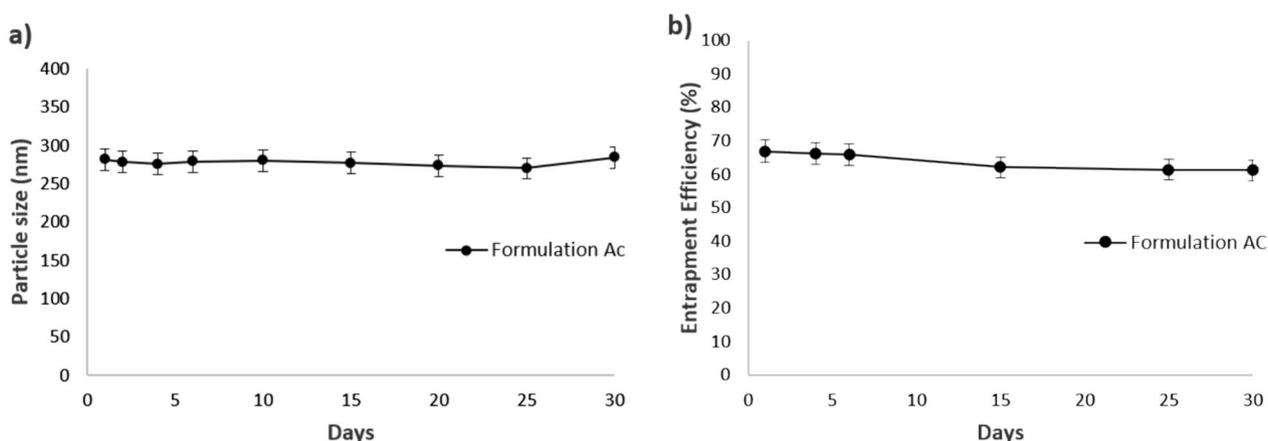

**Fig. 4** Changes in **a** particle size and **b** encapsulation efficiency of berberine chloride-loaded gelatin system in $A_C$ formulation at a storage temperature of 4° C for predetermined time intervals. The trend showed stability of particle size and encapsulation efficiency over 1 month

**Stability of nanodroplets**

Based on the observations made in Fig. 4 a, b, it can be inferred that there were almost no changes in the size and entrapment efficiency of nanodroplets for a period of 1 month at a temperature of 4 °C. This stability period is promising for the development of an effective drug delivery system. However, the acceptability of this 1-month stability period largely depends on the specific requirements and intended use of the drug delivery system. In our previous study (Baghbani et al. 2016), the alginate nanodroplet showed 4 months of stability. Negatively charged nanoparticles like alginate often exhibit greater colloidal stability in aqueous solutions, as they repel each other due to electrostatic repulsion. On the other hand, proteins like gelatin are generally more susceptible to degradation over time compared to polysaccharides which can lead to reduced stability. In this study, we chose material with a positive surface charge in favor of higher cellular uptake and pH-responsive delivery. Positively charged nanoparticles can interact more favorably with negatively charged cell membranes. This electrostatic attraction can enhance cellular uptake, which is beneficial when the goal is to deliver drugs or therapeutic agents inside cells. In addition, gelatin nanodroplets showed pH-responsive behavior that can improve drug delivery to specific tissues.

**Conclusions**

In conclusion, this study successfully developed and optimized gelatin/perfluorohexane (PFH) nanodroplets loaded with berberine chloride as a model drug for dual pH and ultrasound-triggered release. The nanodroplets were fabricated using an emulsion-based technique and the effects of various formulation and process parameters were systematically investigated. An optimal formulation was identified containing 1.67% w/v gelatin, 5 mg berberine chloride, 150 μL PFH, 0.5% v/v Tween 80, and 24,000 rpm homogenization speed. This formulation yielded nanodroplets with a size of 281.7 nm, polydispersity index of 0.24, and encapsulation efficiency of 66.8%. Physicochemical characterization using TEM and FTIR verified the distinct core–shell structure of the nanodroplets and confirmed the successful encapsulation of berberine chloride. In vitro release studies demonstrated that the nanodroplets exhibited pH-responsive behavior, with a higher cumulative release of 27.2% under acidic conditions compared to 15.4% at neutral pH over 24 h. This is attributed to the protonation of the gelatin shell under acidic environments. More importantly, ultrasound stimulation triggered rapid acoustic droplet vaporization and subsequent burst release of the encapsulated drug, especially at a lower frequency of 150 kHz which generated minimal heat. After 10 min of ultrasound exposure, 150 kHz and 1 MHz stimulated the release of 89.4% and 63.9% berberine chloride, respectively. Finally, the optimized nanodroplets displayed excellent stability over 1 month of storage at 4°C.

Overall, this work successfully developed a promising dual-triggered nanocarrier formulation for controlled and targeted release of berberine chloride. The nanodroplets leverage the synergy between pH responsiveness and ultrasound stimulation to achieve enhanced drug release specifically at the target site. This improves the bioavailability and therapeutic efficacy of the drug while reducing systemic side effects. The gelatin/PFH platform presents an efficient, biocompatible, and stable system that can potentially be applied for the delivery of various hydrophobic drugs. Further in vitro cytotoxicity and in vivo studies are warranted to fully evaluate the efficiency of these nanocarriers for practical drug delivery applications. This study provides valuable insights into the rational design and engineering of smart multifunctional nanocarriers for targeted therapy.

### List of abbreviations
kHz      Kilohertz
MHZ      Megahertz
PFH      Perfluorohexane
PFC      Liquid perfluorocarbon
ADV      Acoustic droplet vaporization
EPR      Enhanced permeability and retention
PDI      Polydispersity Index
DLS      Dynamic light scattering
TEM      Transmission electron microscopy
PTA      Alkaline phosphotungstic acid
FTIR      Fourier transform infrared
EE      Encapsulation efficiency
ERA      Effective radiation area
pI      Isoelectric point

### Acknowledgements
Not applicable

### Author contributions
MG was responsible for the synthesis and characterization of the nanodroplets, including the preparation of PFH nanodroplets with a gelatin shell containing berberine chloride, conducted the in vitro drug release studies and contributed significantly to the manuscript preparation and editing. FM designed and supervised the overall study, performed the critical analysis of the data. MG analyzed the results related to the drug release kinetics. She also contributed to the interpretation of the results and wrote the corresponding parts of the manuscript. AB performed the ultrasound exposure experiments and analyzed the impact of ultrasound on drug release. MM-B was responsible for conducting the physicochemical characterization of the nanodroplets, including particle size and zeta potential measurements. MM-D provided expertise in ultrasound technology and guided the application of ultrasound in the experiments. The first draft of the manuscript was written by FM and all authors commented on previous versions of the manuscript. All authors read and approved the final manuscript.

### Funding
The authors declare that no funds, grants, or other support were received during the preparation of this manuscript.

### Availability of data and materials
The datasets and materials used in this research are available upon request. Interested parties may contact the corresponding author (Mahshid Givarian, m.geevarian2@aut.ac.ir) to obtain access to the datasets and materials used in this study.

### Declarations

### Ethics approval and consent to participate
This is an observational study. No ethical approval is required.

### Consent for publication
The manuscript does not contain any person's data in any form (including any individual details, images, or videos).

### Competing Interests
The authors have no relevant financial or non-financial interests to disclose.

### Author details
[1]Biomaterial Group, Faculty of Biomedical Engineering (Center of Excellence), Amirkabir University of Technology, Tehran, Iran. [2]School of Medicine, Mashhad University of Medical Sciences, Mashhad, Iran. [3]Department of Medical Physics, Faculty of Medical Sciences, Tarbiat Modares University, Tehran, Iran. [4]Department of Biomedical and Pharmaceutical Sciences, University of Montana, 32 Campus Drive, Missoula, MT 59812, USA.

## Publisher's Note